\documentclass[12pt,fleqn]{article}
\usepackage{psfig,float}
\usepackage{graphicx}
\usepackage{bm}

\begin{document}

\begin{center}
{\Large{\bf The $\sigma$ meson in
 a nuclear medium through two pion photoproduction}}
\end{center}
\vspace{1cm}

\title{The $\sigma$ meson in a nuclear
 medium through two pion photoproduction}

\begin{center}
{\large{ L. Roca, E. Oset and M.J. Vicente Vacas}}
\end{center}
\vspace{0.4cm}
\begin{center}
{\it Departamento de F\'{\i}sica Te\'orica and IFIC}
\end{center}
\begin{center} 
{\it Centro Mixto Universidad de Valencia-CSIC}
\end{center}
\begin{center}
{\it Institutos de Investigaci\'on de Paterna, Apdo. correos 22085,}
\end{center}
\begin{center}
{\it 46071, Valencia, Spain}       
\end{center}

\begin{abstract}
We present theoretical results for $(\gamma, \pi^0 \pi^0)$ production
 on nucleons and nuclei
in the kinematical region where the scalar isoscalar $\pi \pi$ amplitude is 
influenced by the
$\sigma$ pole. The final state interaction of the pions modified by the nuclear 
medium produces a spectacular shift
of strength of the two pion invariant mass distribution induced by
 the moving of the $\sigma$ pole to lower masses and  widths as the nuclear 
 density increases.

\end{abstract}

The nature of the $\sigma$ meson has been the subject of continuous debate. Its
nature as an ordinary $q \bar{q}$ meson or a $\pi \pi $ resonance has
centered most of the discussion \cite{kyoto}.  The advent of $\chi PT$ has
brought new light into this problem and soon it was suggested 
\cite{Gasser:1991bv,Meissner:1991kz} that the $\sigma$
could not qualify as a genuine meson which would survive in the limit of large
$N_c$. The reason is that
the  $\pi \pi$ interaction in s-wave in the isoscalar sector is strong enough
to generate a resonance through multiple 
scattering of the pions. This seems to be the case, and even in models starting
with a seed of $q \bar{q}$ states, the incorporation of the $\pi \pi$
channels in a unitary approach leads to a large dressing by a pion
cloud which makes negligible the effects of the original $q \bar{q}$ seed 
\cite{torn}.  This idea has been made more quantitative through the introduction
of the unitary extensions of $\chi PT$ ($U \chi PT$) 
\cite{Dobado:1990qm,Dobado:1993ha,OllOsePel,Oller:1999zr}. These works implement
unitarity in coupled channels in an exact form and use the input of the lowest
and second order chiral Lagrangians of \cite{Gasser:1985ux}. The inverse
amplitude method is used in \cite{OllOsePel} and an expansion in powers of
$O(p^2)$ is done for the real part of the inverse of the scattering amplitude,
while in \cite{Oller:1999zr} the dynamics of the second order chiral Lagrangian
is introduced via the explicit use of the exchange of genuine mesons, following
the lines of \cite{Ecker:1989te}, and unitarizing with the N/D method. These
works also justified the success in the scalar sector of the Bethe
Salpeter approach used in \cite{Oller:1997ti,Nieves:2000bx}.  In all these cases
the $\sigma$ meson appears as a pole of the $\pi \pi$ scattering amplitude 
in the second Riemann sheet, even when the
second order Lagrangian, which contains information of the exchange of genuine
mesons according to \cite{Ecker:1989te}, is neglected. 
These unitary models have been tested
with success in many elementary reactions \cite{Oller:2000wa}.

Another point of interest which can help us to understand the nature of the
 $\sigma$ meson is the modification of its properties at finite nuclear density. 
The importance of the medium modification of the $\pi \pi$ interaction in the scalar
sector was suggested in \cite{Schuck:1988jn} where the $\pi \pi$ amplitude in
the medium developed large peaks below the two pion threshold, somehow
indicating that the $\sigma$ pole had moved to much lower energies.  The issue
has been revised and the models have been polished incorporating chiral
constraints \cite{Rapp:1996ir,Aouissat:1995sx,Chiang:1998di} with the result 
that the peaks
disappear at normal density, 
but still much strength is shifted to low energies.

Since present theoretical calculations agree on a sizeable modification in the
nuclear medium of the $\pi\pi$ scattering in the $\sigma$ region, our purpose here 
is to find out its possible experimental signature in a very suited process
 like the $(\gamma,\pi^0 \pi^0)$ reaction in nuclei. Recent
experiments at Mainz \cite{metag}, where preliminary results show a very clear
shift of  strength of the invariant mass distribution of the two pions towards
low invariant masses, seem to indicate that medium effects are indeed large.

In the present paper we shall show how the  $\pi \pi$
interaction in the final state of the $(\gamma, \pi^0 \pi^0)$ reaction on nucleons
enhances the cross section of this reaction and how the medium corrections on
the $\pi \pi$ interaction  in nuclei lead to an appreciable
shift of the strength of the invariant mass distribution towards lower invariant
masses. Although a similar shift has been claimed in the $(\pi, \pi \pi)$ 
reaction in nuclei \cite{Bonutti:1996ij,Bonutti:1998zw,Starostin:2000cb}, the 
fact remains that there are still some discrepancies between these 
experiments and 
that the theoretical calculations \cite{Rapp:1999fx,VicenteVacas:1999xx} 
do not reproduce 
the data. The reason is that the $(\pi, \pi \pi)$ reaction, involving initial 
pions which 
are much distorted in the nucleus, is quite peripheral and the effective 
densities tested
are small. A possible way out to reconcile theory and experiment was suggested in 
\cite{VicenteVacas:1999xx}, showing that the small cross section in 
$\pi^-p\to\pi^+\pi^-n$ at small invariant masses was abnormally small because of a
subtle cancellation of large terms. A medium modification of these terms, through
changes in the $N^*$ properties and others not having to do with the $\pi\pi$
interaction, could offset that cancellation and lead to larger final results.
 The $(\gamma,\pi^0 \pi^0)$ reaction is much better suited to investigate the
 modification of the $\pi\pi$ interaction in the medium
  because the photons are not distorted and one can test larger
nuclear densities.

For the model of the elementary $(\gamma, \pi \pi)$ reaction we follow 
 \cite{Nacher:2001eq} which considers the coupling of the photons to mesons, 
 nucleons, and the resonances
$\Delta(1232)$, $N^*(1440)$, $N^*(1520)$ and $\Delta(1700)$. In the region of
relevance to the present work, $E_\gamma =400-460$~MeV, apart from some
background  terms, the $\Delta$ Kroll Ruderman term, diagram $i)$ of Fig.~1 in  
\cite{Nacher:2001eq}, is of importance and will be dealt with separately 
from the rest. The model of \cite{Nacher:2001eq} relies upon tree level
diagrams. Final state interaction of the $\pi N$ system is accounted for
by means of the explicit use of resonances with their widths. However,
since we do not include explicitly the $\sigma$ resonance, the final state
interaction of the two pions has to be implemented to generate it.

The $\gamma N \to N \pi^0 \pi^0$ amplitude can be decomposed in two parts:
the one that has as final state the combination of pions with isospin I=0,
 first term of the RHS of Eq.~(\ref{eq:T00}),
and the I=2 combination, last two terms of Eq.~(\ref{eq:T00}). 

%\begin{widetext}
\begin{eqnarray}
|\pi^0(1)\pi^0(2)>=
\underbrace{\frac{1}{3}|\pi^0(1)\pi^0(2)+\pi^+(1)\pi^-(2)+\pi^-(1)\pi^+(2)>}
_{\textrm{I=0 part}}\nonumber\\  
%\hspace{0.3cm}
\underbrace{-\frac{1}{3}|\pi^0(1)\pi^0(2)+\pi^+(1)\pi^-(2)+\pi^-(1)\pi^+(2)>
+|\pi^0(1)\pi^0(2)>}_{\textrm{I=2 part}}
\label{eq:T00}
\end{eqnarray}
%\end{widetext}

The interaction of
pions in I=2 in s-wave at these energies is very weak and hence we do not modify
this part of the $\gamma N \to N \pi^0 \pi^0$ amplitude due to the final state
interaction of the pions. However, the I=0 part is strongly modified.  We
have also checked that at the low energies involved here the pions 
come essentially in s-wave.

\begin{figure}[ht!]
\vspace{-0.4cm}
\centerline{\protect\hbox{\psfig{file=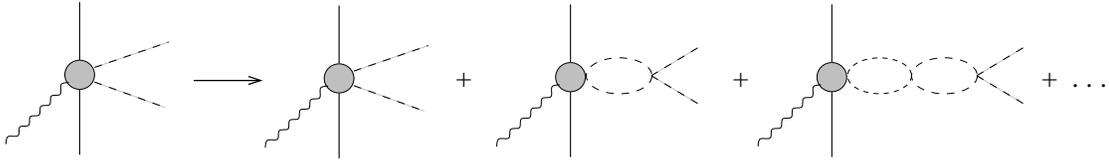,height=2.5cm,width=15.0cm,angle=0}}}
\caption{\small{Diagrammatic series for pion final state interaction in I=0}}
\label{fig:Tgammapipi}
\end{figure}

In ref.  \cite{VicenteVacas:1999xx} the renormalization of the 
$I=0$ $(\pi,\pi \pi)$ amplitude was done by factorizing the on shell 
tree level $\pi N \to \pi \pi N $ and $\pi \pi \to \pi \pi$ amplitudes in the 
loop functions. This was justified in  \cite{Oller:1997ti} for the $\pi \pi$
interaction. The same approach would lead in our case to

\begin{equation}
T_{(\gamma,\pi^0\pi^0)}^{I_{\pi\pi}=0}\to T_{(\gamma,\pi^0\pi^0)}^{I_{\pi\pi}=0}
\left(1+G_{\pi\pi}t_{\pi\pi}^{I=0}(M_I)\right)
\label{eq:GT1}
\end{equation}
where $G_{\pi\pi}$ is the loop function of the two pion propagators, 
which appears in the Bethe Salpeter equation, and $t_{\pi\pi}^{I=0}$ is the
$\pi\pi$ scattering matrix in isospin I=0. 
In order to show clearly how one is lead to this equation, we show in
Fig.~\ref{fig:Tgammapipi} the diagrams involved in the two pion production 
including their final state interaction.

\begin{figure}[ht!]
\vspace{-0.4cm}
\centerline{\protect\hbox{\psfig{file=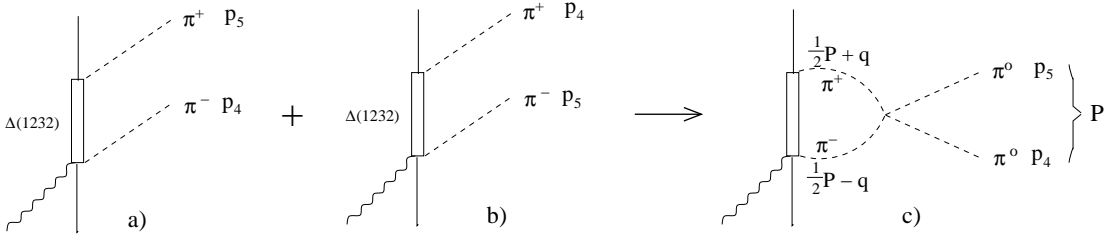,height=3.5cm,width=15.0cm,angle=0}}}
\caption{\small{Diagrammatic series for the $\Delta$ Kroll Ruderman term
with final pions in I=0.}}
\label{fig:FSIKR}
\end{figure}

The two final pions undergo multiple scattering which can be accounted for
by means of the Bethe Salpeter equation, 

\begin{equation}
t=V+VG_{\pi\pi}t
\label{eq:Bethe}
\end{equation}
where, following \cite{Oller:1997ti}, V is given by the lowest order chiral
amplitude for $\pi\pi\to\pi\pi$ in $I=0$
and $G_{\pi\pi}$, the loop function of the two pion propagators, is
regularized by means of a cut off in \cite{Oller:1997ti}, or alternatively with
dimensional regularization in \cite{Oller:1999zr}. In both approaches it was shown
that $V$ factorizes with its on shell value in the Bethe-Salpeter equation of 
Fig.~\ref{fig:Tgammapipi}. Hence, in the Bethe-Salpeter equation the integral
involving $Vt$ and the product of the two pion propagators affects only these
latter two, since $V$ and $t$ factorize outside the integral, thus leading to
Eq.~(\ref{eq:Bethe}) where $VG_{\pi\pi}t$ is the algebraic product of V, the loop 
function of the two propagators, $G_{\pi\pi}$, and the $t$ matrix. Coming back to 
Fig.~\ref{fig:Tgammapipi} it is now clear that in case the vertex from where
the two pions emerge is a contact term with a constant amplitude, the series
implicit in Fig.~\ref{fig:Tgammapipi} is summed as
\begin{eqnarray}
&&T + TG_{\pi\pi}V + TG_{\pi\pi}VG_{\pi\pi}V + \dots=
\nonumber\\%
&&=T [1 + G_{\pi\pi} (V + VG_{\pi\pi}V + \dots )] = T (1 + G_{\pi\pi}t_{\pi\pi})
 \label{eq:Bethe2} 
\end{eqnarray}
which is Eq.~(\ref{eq:GT1}). Now in the model for $( \gamma,2\pi )$ of 
\cite{Nacher:2001eq} there are indeed contact terms as implied before, as well as
other terms involving intermediate nucleon states or resonances. In this
latter case the first loop function in the diagrams of Fig.~\ref{fig:Tgammapipi}
is more complicated since it involves three propagators. Yet,
if the intermediate baryon is far off shell, as is the case for most diagrams, then
the baryon propagator does not change much in the loop function and the
factorization of Eq.~(\ref{eq:GT1}) still holds. There is, however, an exception in
the $\Delta$ Kroll Ruderman term, since as we increase the photon energy we get
closer to the $\Delta$ pole. For this reason, and because its weight is important
at these energies
in the $(\gamma,\pi^+\pi^-)$ amplitude which is needed in Eq.~(\ref{eq:T00}),
we have singled out this
term which we work out in detail below. The I=0 part of the amplitude requires
the combination of the diagrams shown in Fig.~\ref{fig:FSIKR} a) b)

\hspace{-0.8cm} 
and hence the vertex contribution is
(assuming the $\Delta$ propagator, $G_{\Delta}$, the same in both cases)

\begin{equation}
G_{\Delta} (\vec{S}·\vec{p}_5 \ \vec{S}^{\dagger}·\vec{\epsilon}
+ \vec{S}·\vec{p}_4 \ \vec{S}^{\dagger}·\vec{\epsilon})
= G_{\Delta} \vec{S}·(\vec{p}_4+\vec{p}_5) \ \vec{S}^{\dagger}·\vec{\epsilon}
= G_{\Delta} \vec{S}·\vec{P} \ \vec{S}^{\dagger}·\vec{\epsilon}
\end{equation}
where $\vec{S}^{\dagger}$ is the spin transition operator from spin 1/2 to 3/2
and $\vec{\epsilon}$ is the photon polarization.

In the loop function originated from the sum of the terms a) and b) depicted in
Fig.~\ref{fig:FSIKR}, assuming also the $\Delta$ propagator constant, we would 
have the contribution

\begin{eqnarray}
&i \int \frac{d^4 q}{(2\pi)^4}2 \ \vec{S}·(\frac{\vec{P}}{2}+\vec{q})
\ \vec{S}^{\dagger}·\vec{\epsilon} \ G_{\Delta}
 \frac{1}{(\frac{1}{2}P+q)^2-m_{\pi} ^2 + i\epsilon}
\ \frac{1}{(\frac{1}{2}P-q)^2-m_{\pi} ^2 + i\epsilon} \ t_{\pi\pi}=
\nonumber\\
%\hspace{-2cm}
&= G_{\Delta} \ \vec{S}·\vec{P} \ \vec{S}^{\dagger}·\vec{\epsilon} 
\ G_{\pi\pi}(s) \ t_{\pi\pi}(s)
\label{eq:Gdel1}
\end{eqnarray}
since the term proportional to $\vec{S}\vec{q}$ vanishes because of parity reasons,
and then we
see explicitly the factorization of the tree level amplitude. If we keep
explicitly the $\Delta$ propagator in the loop some corrections arise since now we
have the loop function with the $\Delta$ propagator and two pion propagators. 
This loop integral is performed following the steps of \cite{Oset:2000gn}, where
the same loop with three propagators, albeit with only strong vertices, is
evaluated. By performing explicitly the $q^0$ integration in Eq.~(\ref{eq:Gdel1})
we obtain

\begin{eqnarray}
\nonumber
&&\int \frac{d^3q}{(2\pi)^3}
2 \ \vec{S}·(\frac{\vec{P}}{2}+\vec{q})
\ \vec{S}^{\dagger}·\vec{\epsilon} \
\frac{M_{\Delta}}{E_{\Delta}}
\frac{1}{2\omega\omega'} \ \frac{1}{P^0+\omega+\omega'}
\ \frac{1}{P^0-\omega-\omega'+i\epsilon}
\\ \nonumber
&&\cdot\frac{1}{\sqrt{s}-\omega'-E_{\Delta}
+i\frac{\Gamma_{\Delta}(p_{\Delta}^2)}{2}}
\ \frac{1}{\sqrt{s}-\omega-P^0-E_{\Delta}+i\frac{\Gamma_{\Delta}(p_{\Delta}^2)}{2}}
\\ 
&&\cdot\left[(\omega+\omega')(\sqrt{s}-E_{\Delta}-\omega-\omega')-\omega P^0 \right]
\label{eq:Ggorro}
\end{eqnarray}
where $E_{\Delta}=\sqrt{M_{\Delta}\,^2+\vec{p}_{\Delta}\,^2}$,   
$\omega=\sqrt{m_{\pi}^2+(\frac{\vec{P}}{2}+\vec{q}\,)^2}$,  
$\omega'=\sqrt{m_{\pi}^2+(\frac{\vec{P}}{2}-\vec{q}\,)^2}$, 
$p_{\Delta}=\frac{\vec{P}}{2}-\vec{q}$,  
$p_{\Delta}^0=\frac{P^0+\omega+\omega'}{2}$ 
and $\sqrt{s}$ is the CM energy of the initial photon and nucleon.

The improved loop calculation results in a 10 per cent reduction in the cross
section. 
The pion pole term (diagram $j$ of Fig.~1 of ref.~\cite{Nacher:2001eq}) is
essential in the $(\gamma,\pi\pi)$ model to guarantee gauge invariance, but is
numerically negligible at the energies which we have here since it is
proportional to $\vec{P}\,^2$, while the $\Delta$ Kroll-Ruderman term $i)$ is proportional
to $\vec{P}$. Yet, inside the loops, the contribution can be bigger since the
$\gamma\pi\pi$ vertex now is proportional to the loop momentum and not to the
external pion momentum. In addition there is an extra p-wave $\pi N\Delta$ vertex
which also goes like the loop momentum. However, there is also an extra pion
propagator and one also finds cancellations from the poles of the different
propagators. This was done explicitly in \cite{Lee:1998gt} where 
moderate effects from
this pion pole term were obtained. We have performed the appropriate calculation
and found that the loop diagram from the pion pole term is of the order of
10 per cent of the corresponding loop with the $\Delta$ Kroll-Ruderman term.
Hence, we simply take care of it through the factorization approximation.
We also take advantage of this numerical calculation to include 
two extra baryon form factors in the loop, as done in 
\cite{Oset:2000gn}, to account for the $\pi N \Delta$ vertex correction.
This is also done
for the other diagrams since $\pi NN$ vertices are also involved. The inclusion of
the form factors leads to a further reduction of about 15 per cent in the final
cross section.
 Hence, the
actual modification of the amplitude should be
\begin{eqnarray}
\nonumber
T \to 
&& T-T^{00} + \left(T^{00}-T_{KR}^{00} \right) \left(1+G_{\pi\pi}t_{\pi\pi} \right)
+ T_{KR}^{00} + G_{KR} t_{\pi\pi}
\\
&&  =T + \left(T^{00}-T_{KR}^{00} \right) G_{\pi\pi}t_{\pi\pi}
+ G_{KR} t_{\pi\pi}
\label{eq:Tunit}
\end{eqnarray}
where $T$ is the full ($\gamma,\pi^0\pi^0$) amplitude,
$T^{00}$ is the $(I=0,I_{3}=0)$ amplitude for ($\gamma,\pi^0\pi^0$), 
$T_{KR}^{00}$ is the same as $T^{00}$ but calculated only with
 the $\Delta$ Kroll Ruderman term, 
$G_{\pi\pi}$ is the two pions loop including the strong $\pi BB'$ form factors, 
$t_{\pi\pi}$ is the $\pi\pi$ scattering matrix in isospin I=0, 
$G_{KR}$ is the loop of Fig.~\ref{fig:FSIKR} 
including in the integrand the two form factors 
and the $T_{KR}^{00}$ that depends on the momentum in the loops.
(This is, Eq.~(\ref{eq:Ggorro}) including the form factors and the coupling
constants and isospin coefficients of \cite{Nacher:2001eq}).

 After all this is done, we find it technically useful, in order
to account for these elaborate
loop corrections, to still apply the factorization of the
$(\gamma, 2 \pi)$ tree level amplitude of Eq.~(\ref{eq:GT1}) but with a slightly
 modified form
factor included in the $G_{\pi\pi}$ loop function. The cut off of the monopole
form factor is changed from 1~GeV to 625~MeV to implement these changes, 
including also the small effects from the projection in the two pion s-wave channel. 
This procedure is quite
accurate numerically and prevents the numerical task from blowing up when we
perform the calculations in nuclei.

  There is also a small technical detail. One of the terms in our approach
contains the Roper excitation and its posterior decay into two pions in s-wave.
The Lagrangian used is given in \cite{VicenteVacas:1999xx} and the coupling
constant is renormalized, as in \cite{VicenteVacas:1999xx}, in such a way that
when the $\pi \pi$ final state interaction is taken into account the empirical
$N^*$ width is obtained.

The cross section for the nuclear process can be calculated using many
body techniques in a similar way to \cite{Oset:1986qd}, \cite{Salcedo:1988md}.
\begin{figure}
\center \includegraphics[angle=-90,width=7.5cm]{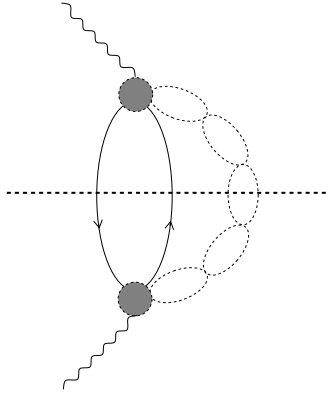}
%\centerline{\protect\hbox{\psfig{file=interference.ps,height=9cm,width=8.0cm,angle=-90}}}
\caption{\small{Selfenergy diagram for the evaluation of the cross section
including the final pions rescattering.}}
\label{fig:self}
\end{figure} 
The method proceeds in two
steps. In the first one the probability per unit length of a $(\gamma,2\pi)$
process in nuclear matter is evaluated from the imaginary part of the photon
selfenergy diagram of Fig.~\ref{fig:self}
corresponding to the cut of the horizontal line.
By using the local density
approximation and, hence, substituting the density of nuclear matter by the
empirical density of the nucleus $\rho(\vec{r})$ at a point $\vec{r}$, this
provides the reaction probability in this point of the nucleus. 
The second step requires to follow the individual
particles produced through the nucleus in an inclusive process where the nucleus
can be excited to any state. This second step is done by
using semiclassical methods in which the pions produced follow classical
trajectories and are allowed to undergo quasielastic
collisions or pion absorption according to probabilities calculated previously
using many body techniques. As shown in \cite{Salcedo:1988md}, this procedure was
very successful in describing pion nucleus phenomenology at intermediate energies.

 With these ingredients the nuclear cross section is given by
  
\begin{eqnarray}
\sigma=&&\frac{\pi}{k}\int d^3\vec{r}\int\frac{d^{3}\vec{p}}{(2\pi)^3}
\int\frac{d^{3}\vec{q_1}}{(2\pi)^3} 
\int\frac{d^{3}\vec{q_2}}{(2\pi)^3}\frac{1}{2\omega(\vec{q_1})}\frac{1}{2\omega(\vec{q_2})}
\nonumber\\%
&&\cdot
\sum_{s_i,s_f}\overline{\sum_{pol}}\mid T\mid^{2} n(\vec{p}) \ [1-n(\vec{k}+\vec{p}
 -\vec{q_1}-\vec{q_2})]
\nonumber\\
&&\cdot \ \delta(k^{0}+E(\vec{p})-\omega(\vec{q_1})-\omega(\vec{q_2})-E(\vec{k}
+\vec{p}-\vec{q_1}-\vec{q_2}))\nonumber\\
&&\cdot \ F_1(\vec{r},\vec{q_1}) \ F_2(\vec{r},\vec{q_2})
 \label{eq:sigma2}
\end{eqnarray}
where $n(\vec{p})$ is the occupation number for a density $\rho(\vec{r})$.
The factors $F_i(\vec{r},\vec{q_i})$  take into account the distortion of
 the final pions in their way out through the nucleus and are given by
 
 \begin{eqnarray}\hspace{0cm}
F_i(\vec{r},\vec{q_i})=exp\left[\int_{\vec{r}}^{\infty}dl_i \ \frac{1}{q_i}Im
\Pi (\vec{r}_i) \right]
\label{eq:Feikonal}
\\ \nonumber & &\vspace{0.4cm}\hspace{-5.5cm}
\vec{r_i}=\vec{r}+l_i \ \vec{q_i}/\mid \vec{q_i}\mid
\end{eqnarray}
where $\Pi$ is the pion selfenergy, taken from \cite{Nieves:1993ye},
where the interaction of low energy 
pions with nuclear matter was studied. This potential has been tested against 
the different pionic reaction cross sections, elastic, quasielastic
and absorption. The imaginary part of the potential is split into a part that
accounts for the probability of quasielastic collisions and another one which
accounts for the pion absorption. As we shall see, the probability
that there is loss of pion flux through pion absorption at low energies is 
 larger than through quasielastic collisions. One of the reasons is the Pauli
blocking of the occupied states.

 When we renormalize the I=0 amplitude to account for the pion final state
interaction, we change $G_{\pi\pi}$ and $t_{\pi\pi}^{I=0}$ by their corresponding results
in nuclear matter \cite{Chiang:1998di} evaluated at the local density of the 
point $\vec{r}$ in the integral of Eq.~(\ref{eq:sigma2}).
We do not include here the direct coupling of the two pions to the 
$N^*(1440)h$
which was found extremely small in \cite{Oset:2000ev}. We do not include
either the direct coupling of the two pions to a $p-h$ excitation. The s-wave
would proceed only through the tiny isoscalar $\pi N\to \pi N$ interaction, and
the p-wave part, weak in principle at the relatively small momenta of the pions,
would be only relevant at excitation energies of about 30-50~MeV, far below the
region of interest to us in the present problem. On the other hand, one should
note that other alternatives have been proposed to study the modification of the
isoscalar $\pi\pi$ interaction in the nuclear medium based on the reduction of the
pion decay coupling constant in the medium, which is tied to the dropping of the
quark condensate via the Gell-Mann--Oakes--Renner relation
\cite{Hatsuda:1999kd,Aouissat:2000ss},
or a combination of this effect and the dressing of the pion in the medium 
\cite{Davesne:2000qj}. However, in the approach which we follow, one must be
cautious not to include the dropping of $f$ on top of the many body corrections
done. Indeed, the change of $f$ as a function of density in
\cite{Hatsuda:1999kd,Hatsuda:2001da} and related works, obtained through 
the GOR relation, is linked to the renormalization of the time component of the
axial current in nuclei, but in standard many body theory these currents are
renormalized using the same lagrangians as in the evaluation of the pion
selfenergy \cite{Kubodera:1978wr,khanna:78,oset_rho,Kirsch}.
 Therefore, one should not modify $f$ in these latter approaches to
avoid double counting if using an explicit perturbative calculation with
effective chiral lagrangians. Actually, a large change of $f$ at normal nuclear
density would be difficult to accommodate to the quite well known pion nucleus
phenomenology (pionic atoms, pion absorption, etc). Concerning this point it is
worth noting the discussion in \cite{Harada:1999zj,Harada:2001kb,Harada:2001at} about the difference between the $f(\rho)$ constant related
to the quark condensate and the $f$ used in perturbative calculations with chiral
lagrangians. These differences were also stressed in \cite{Garcia-Recio:2002se},
where it was also shown that taking the $f(\rho)$ coupling related to the quark
condensate in the many body evaluation of the pion nucleus optical potential led
to unacceptably large widths of pionic atoms.

We have also used the complex $\Delta$ selfenergy from \cite{Oset:1987re} to 
dress the
$\Delta$ propagator. In addition to the proper real part of the selfenergy in 
\cite{Oset:1987re} we add the effective contribution to the selfenergy 
$4/9(f^{*}/\mu)^2g'\rho$ coming from the iterated
$\Delta$h excitation driven by de Landau Migdal interaction 
\cite{Carrasco:1992vq,Oset:1987re}.

\begin{figure}
\center \includegraphics[angle=-90,width=8.5cm]{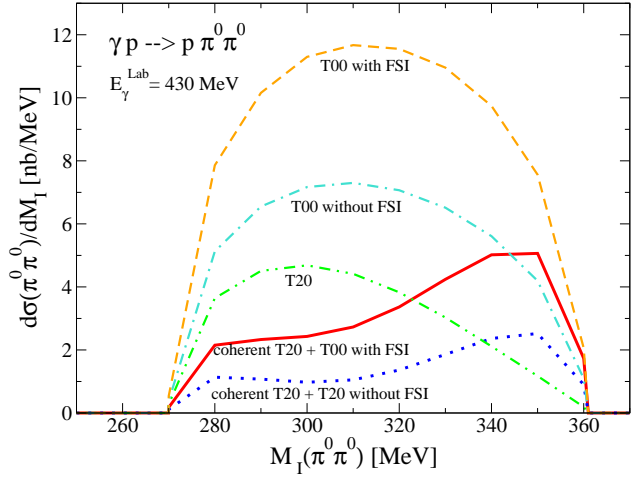}
%\centerline{\protect\hbox{\psfig{file=interference.ps,height=9cm,width=8.0cm,angle=-90}}}
\caption{\small{Contributions of the different isospin channels to the $2\pi$
 mass distribution.}}
\label{fig:interf}
\end{figure} 

   In Fig.~\ref{fig:interf} we show the results of the invariant mass 
distribution of the two pions for the $\gamma p \to \pi^0 \pi^0 p$ reaction.
 We can see the contribution of the I=0 part alone (T00),
the I=2 part alone (T20) and the coherent sum of the two, both with and without
renormalization (FSI) of the I=0 amplitude.  The
renormalization of the I=0 amplitude has important effects, nearly doubling
the cross section. This is reminiscent of the similar enhancement found from
chiral loops at threshold in \cite{Bernard:1994ds}.
 When  we sum coherently the I=0 and I=2 amplitudes the shape of the 
 distribution exhibits a double hump, one at low
invariant masses and the other one at the high mass part of the spectrum.  This
shape is corroborated by the preliminary experimental results of \cite{metag}. 

\begin{figure}
\center \includegraphics[angle=-90,width=8.5cm]{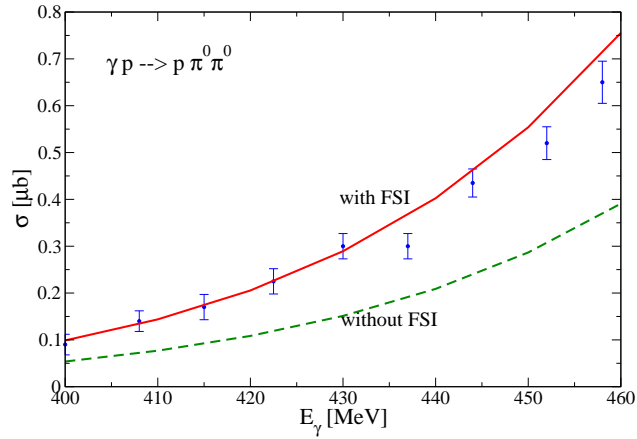}
%\centerline{\protect\hbox{\psfig{file=X_400_460.ps,height=9cm,width=7.0cm,angle=-90}}}
\caption{\small{Total cross section for $\gamma p \to \pi^0 \pi^0 p$ with and without pion
final state interaction. Experimental data from ref. \cite{Wolf:2000qt}.}}
\label{fig:X-free}
\end{figure}

  The integrated cross section compared with experiment can be seen in 
  Fig.~\ref{fig:X-free}
where we can appreciate that the inclusion of final state interaction
 improves the agreement with the data.  The amplitude for the $(\gamma,
\pi^0 \pi^0)$ reaction on neutrons is calculated along the same lines and 
leads to a similar mass distribution as that of Fig.~\ref{fig:interf} albeit 
with a little smaller cross section.

\begin{figure}
\includegraphics[angle=0,width=15cm]{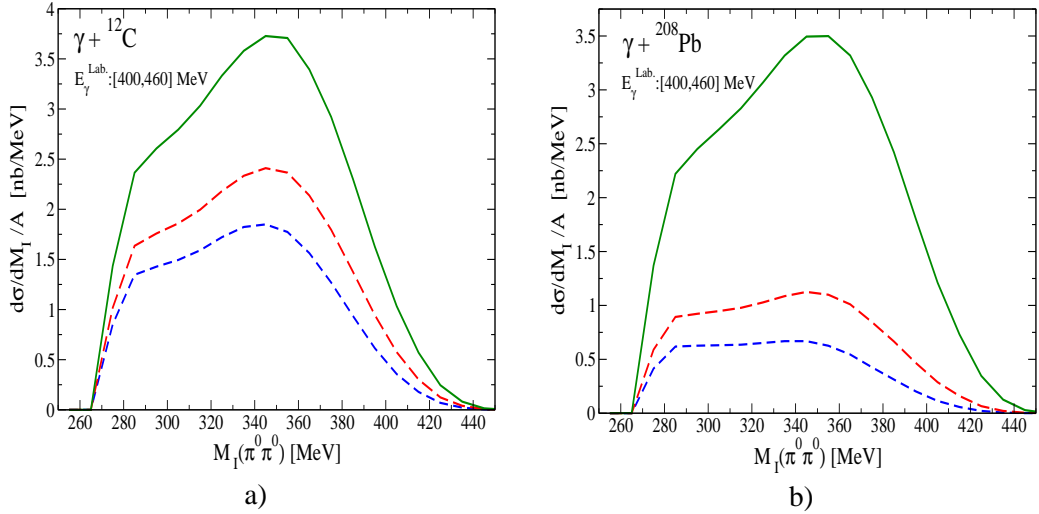}
%\centerline{\protect\hbox{\psfig{file=X_400_460.ps,height=9cm,width=7.0cm,angle=-90}}}
\caption{\small{Two pion invariant mass distribution for $2\pi^0$ photoproduction 
in $^{12}C$ and $^{208}Pb$. All three curves are calculated using the $2\pi$ final state interaction at
 0 density but they differ in the final pion distortion:
continuous line: without absorption nor quasielastic scattering. Long dashed line: only final
pions absorption. Short dashed line: final pions absorption and quasielastic scattering.}}
\label{fig:rho0}
\end{figure}

In Fig.~\ref{fig:rho0} a) and b) we show the cross section for $^{12}C$ and $^{208}Pb$
assuming the
FSI of the pions in the free case $(\rho=0)$, and we show the results without pion
absorption or quasielastic steps, with  pion absorption alone, and a third
 curve which corresponds to the case
where the pions which undergo quasielastic collisions together with those absorbed 
are eliminated from the
outgoing pion flux. This is actually not the case but the comparison of the two
lower curves gives us a measure of the amount of quasielastic collisions undergone
by the pions in their way out from the production point. The figures show that more
pions undergo absorption than quasielastic collisions.
The larger part of the quasielastic collisions would not change the charge
of the pions, only their energy and
 momentum would be changed. In this case the two $\pi^0$ would still be there
 and their invariant mass would be somewhat changed, sometimes leading to larger
 two pion invariant masses and other to smaller ones. Hence, as an average the
 invariant mass distribution of the two pions should not be much modified by these
 quasielastic collisions.
  In other cases there could
 be change of charge and then we would not have two $\pi^0$ in the final state.
 However, this could also be compensated by having originally a $\pi^+\pi^0$
  production
 followed by a collision of the $\pi^+$ with charge exchange. 
The distortion
 of the final pions has been done simply by using the imaginary part of the optical
 potential and leads to a reduction of the cross section.  The results for the
  case of removal of only the pions which are absorbed are
 obtained  by putting in the imaginary part of the pion selfenergy, $\Pi$, in
 Eq.~(\ref{eq:Feikonal}),
 the part which comes from the absorption and omitting the one that comes from
 quasielastic, which have been separated in \cite{Nieves:1993ye}. Additional effects
 from the real part of the potential can be taken into account
 following the lines of \cite{GomezTejedor:1995yr} by introducing
 the real part of the pion selfenergy in the pion propagators
 cut by the horizontal line in Fig.~\ref{fig:self} when one evaluates
 the photon selfenergy with a $ph$ and two pions in the intermediate state. We find a moderate increase
 of the cross section by an amount of 20 percent in  $^{12}C$ from this effect
when the $\pi\pi$ 
interaction in the medium is considered in addition, but again it does
 not modify at all the shape of the invariant mass distribution.  All these
 things discussed, we can conclude that while there are uncertainties of
 about 20 percent in $^{12}C$, and a bit more in $^{208}Pb$, in the size of the 
 total cross section, the shape of the mass
 distribution is still determined basically by the implementation of the $\pi \pi$ 
 interaction in the medium.
 
\begin{figure}
\center \includegraphics[angle=-90,width=8.5cm]{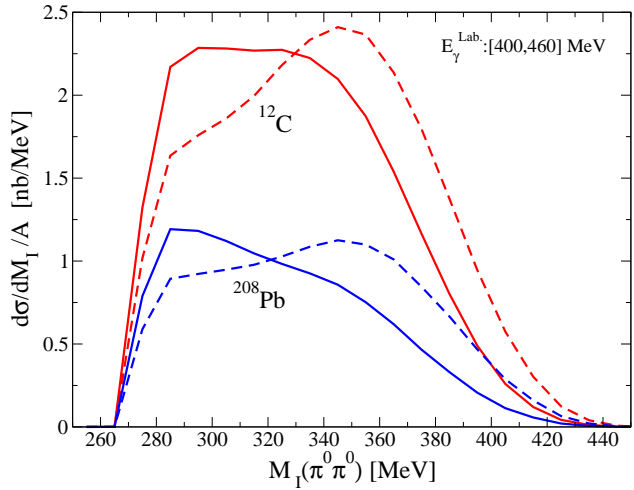}
%\centerline{\protect\hbox{\psfig{file=C12_Pb208.ps,height=8.5cm,width=8cm,angle=-90}}}
\caption{\small{Two pion invariant mass distribution for $2\pi^0$ photoproduction 
in $^{12}C$ and $^{208}Pb$. Continuous lines: Using the in medium final $\pi\pi$ interaction. Dashed
lines: 
using the final $\pi\pi$ interaction in free space.}}
\label{fig:C12}
\end{figure} 

In Fig.~\ref{fig:C12} we can see the results for the invariant mass distribution
of the two pions for $^{12}C$ and $^{208}Pb$ including the absorption of the final
pions.  The difference between the 
solid and dashed curves is the use of the in
medium $\pi \pi$ scattering and $G$ function instead of the free ones, which
we take from \cite{Chiang:1998di}.  As one can see in the figure, there 
is an appreciable shift of strength to the low
invariant mass region due to the in medium  $\pi \pi$  interaction.   This
shift is remarkably similar to the one found in the preliminary measurements of
\cite{metag}.

These results show a clear signature of the modified $\pi\pi$
interaction in the medium.  The fact that the photons are not distorted has
certainly an advantage over the pion induced reactions and allows one to see 
inner parts of the nucleus. In this sense it is worth noting that from our
calculations we can determine the average nuclear density felt by the reaction
which turns out to be 35 per cent and 65 per cent of the normal nuclear density
 for $^{12}C$ and $^{208}Pb$ respectively.

Although we have been discussing the $\pi \pi$ interaction in the nuclear
medium it is clear that we can relate it to the modification of the $\sigma$
in the medium. We have mentioned that the reason for the shift of strength to
lower invariant masses in the mass distribution is due to the accumulated
strength in the scalar isoscalar $\pi \pi$ amplitude in the medium. Yet, this
strength is mostly governed by the presence of the $\sigma$ pole and there have
been works suggesting that the $\sigma$ should move to smaller masses and widths
when embedded in the nucleus 
\cite{Hatsuda:1999kd,Davesne:2000qj,Bernard:1987im,Wambach:2001sh,Aouissat:2000ss}. We should stress that in 
refs.~\cite{Hatsuda:1999kd,Davesne:2000qj,Wambach:2001sh,Aouissat:2000ss} effects
of chiral symmetry restoration from the dropping of the condensate
and/or the change in the bare $\sigma$ mass are considered. In our approach which, as we
mentioned, relies upon a conventional many body expansion based on standard
chiral lagrangians, where the $\sigma$ in the free space or in the nucleus is
generated dynamically, we can look at the $\sigma$ poles and see their evolution
with the nuclear density. A detailed
study of the poles in the complex plane for the $\pi\pi$
interaction in the nuclear medium within this approach 
is now available in \cite{Vacas:2002se}. In this work it is indeed found
that the pole position of the $\sigma$ moves to smaller energies as the density
increases and the width is also reduced. The present results, when contrasted by
the definitive data, if they confirm the preliminary ones of \cite{metag}, 
should represent an evidence of this interesting
phenomenon which would come to strengthen once more the nature of the $\sigma$
meson as dynamically generated by the multiple scattering of the pions through
the underlying chiral dynamics.

\subsection*{Acknowledgments}

One of us, L.R. acknowledges support from the Consejo Superior de
Investigaciones Cient\'{\i}ficas.
This work is also
partly supported by DGICYT contract number BFM2000-1326, and the 
E.U. EURODAPHNE network contract no. ERBFMRX-CT98-0169.

\end{document}